\documentclass[12pt]{article}
\usepackage[cp1251]{inputenc}

\title{\bf $Js$ the $a_0(1450)$ a candidate for the lowest $q\overline{q}~ ^3P_0$
state ?}
\author{A.M.Badalian\\
Institute of Theoretical and Experimental Physics,\\
B.Cheremushkinskaya 25, 117218 Moscow,Russia}
\date{}
\begin{document}

\maketitle

\newcommand{\be}{\begin{equation}}
\newcommand{\ee}{\end{equation}}

\def\la{\mathrel{\mathpalette\fun <}}
\def\ga{\mathrel{\mathpalette\fun >}}
\def\fun#1#2{\lower3.6pt\vbox{\baselineskip0pt\lineskip.9pt
\ialign{$\mathsurround=0pt#1\hfil##\hfil$\crcr#2\crcr\sim\crcr}}}

\begin{abstract}

For the $a_0(1450)$, considered as the $q\overline{q}~1~^3P_0$
state, ``experimental'' tensor splitting, $c_{exp}=(-150\pm 40$)
MeV, appears to be in contradiction with conventional theory of
fine structure. There is no such discrepancy if the $a_0(980)$
belongs to the $1~^3P_J~q\overline{q}$ multiplet. The hadronic
shift of the $a_0(980)$ is shown to be strongly dependent on the
value of  the strong coupling in spin-dependent interaction.

\end{abstract}

\section{Two possible candidates for the ground $^3P_0$ state}

At present a few low-lying $P$-wave light mesons, both in
isovector and isoscalar channels, were experimentally observed
\cite{1,2}, however, theoretical identification of these states
faces with serious  difficulties, first of all, because the
$^3P_0$ member of the lowest $^3P_J$ multiplet is still not
identified in unambiguous  way. Here we shall discuss only more
simple isovector multiplet.

As it is well known the $a_0(980)$ couples strongly to $\eta\pi$
and  $K\overline{K}$ channel and therefore  this state was
suggested to be interpreted as a multiquark meson \cite{3} or
 $K\overline{K}$ molecula \cite{4}. Then under such
assumptions the  $1~^3P_0$ $q\overline{q}$ state is to be
identified with the $a_0(1450)$ meson \cite{5}. There exist also
many arguments in favour to consider the $a_0(980)$  as  the
ground $q\overline{q}~^3P_0$ state, which however  has  large
$K\overline{K}$ component in its wave function \cite{6}.

We shall discuss here both possibilities  performing  the  fine
structure (FS) analysis of the lowest $q\overline{q}~^3P_J$
multiplet. It is evident that a identification  of the members of
this multiplet should be in accord with values and sign  of FS
parameters. We shall show here that in two cases  depending on
whether the $a_0(980)$ or the $a_0(1450)$ is
$q\overline{q}~1~^3P_0$ state, the drastic difference in value and
even sign  of tensor splitting takes place. In particular, if the
$a_0(1450)$ is identified  as the $q\overline{q}~1~^3P_0$ state
then large magnitude and negative sign of tensor  splitting
appears to be in contradiction with conventional QCD theory of the
FS. Also in this case the center of gravity of the $1~^3P_J$
multiplet $M_{cog}$ has large value and large shift with respect
to the mass of the $b_1(1235)$ meson  (see the discussion in Ref.
\cite{7}). In other case if the $a_0(980)$ is the  $1~^3P_0$
state, FS splittings  and $M_{cog}$ appear to be in agreement with
theoretical picture and the value of the hadronic shift is
strongly correlated with the strong coupling in spin-dependent
interaction. For large $\alpha_s(1-loop)= 0.53$ the hadronic shift
appears to be equal zero and the masses of the $a_0, a_1, a_2$
mesons just coincide with their experimental values. For the
$\alpha_s(2-loop)=0.43$ the hadronic shift about 100 MeV is
obtained.

\section{Fine-structure splittings}

For any $n^3P_J$ multiplet the FS parameters: spin-orbit (SO)
splitting $a(nP)$ and tensor splitting $c(nP)$ can be expressed
through the masses $M_J(J=0,1,2)$ of the members of the multiplet
in  the standard way \cite{8}:
\be
a=\frac{5}{12} M_2 - \frac{1}{4}M_1 - \frac{1}{6} M_0,\label{1}\ee
\be
c=\frac{5}{6}M_1 - \frac{5}{18} M_2 - \frac{5}{9}M_0.\ee where the
experimental values of $M_1$ and $M_2$ for the $a_1$ and $a_2$
mesons are well known \cite{1},
\be
M_1=(1230\pm 40)~~\mbox{MeV},~~~M_2=(1318\pm
0.6)~~\mbox{MeV},\label{3}\ee while the  $1~^3P_0$ state can  be
identified either with the $a_0(1450)$ with the mass $M_{0A}$
\cite{1},
\be
\mbox{Case A:}~~~M_{0A} = (1452 \pm 8)~\mbox{MeV} \label{4}\ee or
with the $a_0(980)$ having the mass $M_{0B}$ \cite{1}:
\be
\mbox{Case B:}~~~M_{0B} = (984.8 \pm 1.4)~\mbox{MeV}\label{5}\ee
Then from Eqs.(\ref{1})-(\ref{4}) in case A the following
``experimental'' values of SO and tensor splittings  can be
extracted,
\be
a_A(exp)=(-0.3\pm 11.6)~\mbox{MeV}\label{8}\ee
\be
c_A(exp) = (-148\pm 38)~\mbox{MeV}, \label{9}\ee i.e SO splitting
is compatible with zero or even small negative value while tensor
splitting (7) is always negative and has large magnitude:
\be
-186~\mbox{MeV} \leq c_A(exp) \leq -110~\mbox{MeV}\label{ab}\ee
Note that  large experimental error in $c_A$ (exp) (7) comes from
the error in the $a_1(1260)$ mass (\ref{3}).

In case B, if the $a_0(980)$ is the ground $q\overline{q}~^3P_0$
state, the situation is complicated by strong coupling of the
$q\overline{q}$ channel with other hadronic channels,  $\eta\pi$
and $K\overline{K}$, and for our analysis it is convenient to
introduce the mass $\tilde{M}_0$ of the $1~^3P_0$ state in
one-channel approximation. Then
\be
M_{0B} (exp) = 985 ~\mbox{MeV} = \tilde{M}_0 -
\Delta_{had},\label{ac}\ee where the hadronic  shift
$\Delta_{had}$ is an unknown value while  the mass $\tilde{M}_0$
can be calculated in different theoretical approaches, e.g. in the
paper \cite{9} of Godfrey, Isgur $\tilde{M}_0(GJ)=1090$ MeV which
corresponds  to $\Delta_{had}(GJ)=105$ MeV. In QCD string approach
our calculations give close number for the choice of
$\alpha_s(FS)=0.42.$

It is of interest to note that if the $a_0(980)$ is the
$q\overline{q}$ $1~^3P_0$ state then both SO and tensor splittings
are positive and the situation looks like in charmonium. For the
$\chi_c(1P)$ mesons $a_{exp}(c\overline{c})$ and
$c_{exp}(c\overline{c})$ are both positive with tensor splitting
being by 13\% larger than SO one \cite{10}:
$$ a_{exp}(c\overline{c})=(34.6\pm 0.2)~\mbox{MeV}$$
\be
c_{exp}(c\overline{c})=(39.1\pm 0.6)~\mbox{MeV} \label{ad}\ee

\section{Theory of the $P$-wave fine structure}

Spin--dependent interaction in light $n\overline{n}$ mesons for
the $L$-wave multiplets with $L\not=0$ can be considered as a
perturbation since FS splittings are experimentally small with
compare to the meson masses. It is also assumed that the
short-range one-gluon-exchange (OGE) gives the dominant
contribution to perturbative potentials, in particular, to SO
potential $V_{SO}^P(r)$ and tensor potential $V^P_T(r)$ \cite{11}.
As a result the matrix elements (m.e.) $a_P=\langle
V^P_{SO}\rangle $ and $c_P=\langle V^P_T\rangle$in light mesons
are defined as the  first order terms in $\alpha_s$:
$$a_P(nP)=a_P^{(1)} = \frac{2\overline{\alpha}_s}{m^2_n} \langle
r^{-3}\rangle_{nP},$$ \be c_P(nP) = c^{(1)}_P =
\frac{2}{3}a^{(1)}_P=\frac{4}{3}\frac{\overline{\alpha}_s}{m^2_n}
\langle r^{-3}\rangle_{nP},\label{14}\ee These expressions contain
the constituent  mass $m_n$ of a light quark, which was  supposed
to be fixed and just the same for all states with different
quantum numbers $nL$; $m_n\cong 300 \div 350$ MeV is usually taken
[11].

More detailed analysis of spin-dependent potentials for light
mesons, both perturbative and nonperturbative (NP), was done in
Ref. \cite{12} where Feynman-Schwinger-Fock (FSF) representation
of light meson Green's function was used. Keeping only bilocal
correlators of the fields, the spin--dependent potentials were
expressed through these correlators  (see Appendix) from which
perturbative SO and tensor splittings are given by the
expressions:
\be
a^{(1)}_P(nL) = \frac{2\overline{\alpha}_s}{\mu^2(nL)} \langle
r^{-3}\rangle_{nL},~~~c^{(1)}_P(nL) =
\frac{2}{3}a^{(1)}_P(nL),\label{15}\ee and NP contributions are
\be
a_{NP} (nL) = -\frac{\sigma}{2\mu^2(nL)r}~\langle
r^{-1}\rangle_{nL}\label{16}\ee $$c_{NP}(nL)~\mbox{is compatible
with zero}$$ It is important to stress that in FSF representation
the expansion in inverse quark masses as in heavy quarkonia is not
used, and the constituent mass $\mu(nL)$ in Eqs.(12),(13) is
defined  by the average over the kinetic energy term of the string
Hamiltonian \cite{13,14}. This ``constituent'', or dynamical, mass
$\mu(nL)$ for light quark with current mass $m=0$ appears to be
\be
\mu(nL) = \langle\sqrt{p^2}\rangle_{nL} = \frac{1}{2}\sigma
\langle r \rangle_{nL}~~\mbox{for linear}~~\sigma
r~~\mbox{potential}\ee and can be
 expressed only through  string tension $\sigma$ and
the universal number.

In contrast  to the  constant  mass $m_n$ in Eqs.(\ref{14}), the
mass $\mu(nL)$ \underline{depends}  on the quantum numbers $n,L$
of a given state and  it  is increasing with growing $n$ and $L$.

In light mesons which have large radii linear static potential
$\sigma r$ dominates for all states with exception of the 1S and
1P states where Coulomb term turns out to be important  \cite{14}.

The choice of the string tension $\sigma$ and coupling
$\alpha_{st}$ in static potential can be fixed from the slope and
the intercept of leading Regge $L$ -- trajectory which is defined
for spin-averaged masses $M_{cog}(L)$  for the multiplets $1L$. As
shown in Ref. \cite{14} the experimental values of $M_{cog}(L)$
put  strong restrictions on  $\sigma$ and also strong coupling
$\alpha_{st}$: $\sigma=(0.185 \pm 0.005)$ GeV$^2$, $\alpha_{st}\la
0.40$. Here we present  our calculations  for linear $\sigma r$
potential with $\sigma=0.18$ GeV$^2$ and also for linear plus
Coulomb potential with
\be
\alpha_{st}=0.42,~~~ \sigma = 0.18~\mbox{GeV}^2\ee and take into
account that the experimental Regge $L$-trajectory \cite{14}:
$M^2_{cog}(L)=(1.60\pm 0.04)$ GeV$^2$ gives
\be
M_{cog}(1P) = 1260\pm 10~\mbox{MeV} \ee  Linear  potential with
$\sigma=0.18$ GeV$^2$
 fits very well the orbital excitations with $L\geq 2$. From Ref.
\cite{14} the m.e. can be taken,
\be
\mu_0(1P)=448~\mbox{MeV},~~~\langle r^{-1}\rangle_{1P} =
0.236~\mbox{GeV},~~~ \langle r^{-3}\rangle_{1P} =
0.0264~\mbox{GeV}^3.\ee and the constituent mass for the 1S state,
\be
\mu(1S) = 335 ~\mbox{MeV},\label{21}\ee   coincides with the
conventional value $m_n\approx 300 \div 350$ MeV, usually used in
potential models. For the 1P state calculated m.e. for Cornell
potential are
\be
\mu(1P) = 486~\mbox{MeV},~\langle r^{-1}\rangle_{1P} =
0.260~\mbox{MeV},~\langle r^3\rangle = 0.0394~\mbox{GeV}^3
\label{22}\ee so here $\langle r^{-3}\rangle_{1P}$ turns out to be
by 50\% larger than for linear potential.

NP contributions to SO and tensor potentials can be correctly
defined in bilocal approximation (see Appendix)  when the Thomas
term dominates in NP SO splitting,
\be
a_{NP} (nP) = -\frac{\sigma}{2\mu^2(nP)} \langle
r^{-1}\rangle_{nP}, \label{26}\ee while NP tensor splitting
$c_{NP}(1P)$ for light mesons (as well as for heavy mesons) is
compatible with zero:
\be
0\leq c_{NP} (1P) < 5~\mbox{MeV} \label{27}\ee and therefore can
be neglected in our later analysis. This result follows from the
fact, established in lattice QCD, that vacuum correlator $D_1$,
which defines $c_{NP}$, is  small \cite{15},[16].

Thus  NP contribution is  present only in SO splitting so that
total SO splitting,
\be a=a_P +a_{NP} = \frac{2\overline{\alpha}_s}{\mu^2(1P)} \langle
r^{-3}\rangle_{1P} - \frac{\sigma}{2\mu^2(1P)} \langle
r^{-1}\rangle_{1P}\ee Due to the negative Thomas precession term a
cancellation  of perturbative and NP terms in $a(total)$ takes
place and in principle $a(total)$ could be small or even negative
number.

In contrast to that, NP  tensor splitting is small and positive,
so that total tensor splitting appears to be always positive,
\be
c=c_P = \frac{4~\overline{\alpha}_s}{3\mu^2(1P)} \langle
r^{-3}\rangle_{1P} \label{29}\ee
 Note that in the $\chi_c$ mesons the correction  of order $\overline{\alpha}_s^2$
to $c_P$  was obtained to be also positive and small   \cite{10}.

\section{Remarks about strong coupling $\alpha_s$}

In heavy quarkonia  strong coupling $\alpha_s(\mu_{ren})$ at the
renormalization scale $\mu_{ren}$ can explicitly be extracted from
experimental values of FS splittings due to rather simple,
renorm-invariant relation between $\alpha_s(\mu_{ren})$ and the
combination $\eta=\frac{3}{2} c - a$ \cite{10,15}. In light mesons
one-loop perturbative  corrections ($\alpha^2_s$ order) are still
not calculated  and OGE contributions   (12) with a fitting
$\overline{\alpha}_s$ are assumed to be dominant.

However, at present we know some useful features of $\alpha_s$.

1. The strong coupling freezes at large distances and therefore
$\overline{\alpha}_s$ in OGE terms have to be less, or equal, the
critical value $\alpha_{cr}$ [17],\cite{9}.

2. The critical value $\alpha_{cr}$ was  calculated in background
field theory [17] and obtained $\alpha_{cr}(r)$ have appeared to
be in good agreement with lattice measurement of static potential
in quenched  approximation [18]. For QCD constant
$\Lambda^{(0)}_{QCD}=(385\pm 30)$ MeV, defined in lattice
calculations [19], in Ref.[17] calculated $\alpha_{cr}$ is
\be
\alpha_{cr}(1-loop)=0.59;~~~\alpha_{cr}(2-loop) =
0.43^{+0.05}_{-0.04}~~(n_f=0) \ee

3. The characteristic  size of FS interaction in the $P$-waves,
$R_{FS}$, can be defined as
\be
R_{FS} = \Biggl [ \sqrt[3]{\langle r^{-3}\rangle_{1P}}\Biggr
]^{-1} \cong 0.60~\mbox{fm}\ee From the study of $\alpha_s(r)$ in
coordinate space [17] it was observed that  at distances $r\sim
0.6$ fm strong coupling $\alpha_s(r)$ is already close to the
critical value (24) being only  by about 10\% smaller. Therefore
one can expect that $\overline{\alpha}_s$ in OGE splittings
(22),(23) has to be equal
\be
\overline{\alpha}_s(R_{FS}) = 0.41\pm 0.02\ee

4. The size $R_{FS}$ (25) remarkably coincides with the radius of
the 1P $c\overline{c}$ and also of the 2P $b\overline{b}$ state
which both are about $0.60$ fm. For them the extracted from
experiment  strong coupling is
\be
\alpha_{exp}(c\overline{c},1P) = 0.38 \pm 0.03 (exp), \ee i.e.
this number is very close to that in  Eq.(26). To check a
sensitivity of FS splitting to the choice of $\overline{\alpha_s}$
here in our calculations  we shall take

\be \overline{\alpha}_s(2-loop) =43,~~  \overline{\alpha}_s
(1-loop) = 0.53\cong0.9 \alpha_{cr} \ee

\section{The $a_0(1450)$ is the $q\overline{q}$ $1~^3P_0$ state}

We start with $NP$ contribution to SO splitting $a_{NP}$ and for
$\sigma = 0.18$ GeV$^2$ $(\mu(1P)=448$ MeV)
\be a_{NP} = -106~\mbox{MeV} \label{33}\ee It can be shown that
$a_{NP}$ is weakly dependent on the choice of the parameters of
static interaction varying in  the range 99 MeV -- 106 MeV.

From Eq.(6) $a_A(exp)$ is compatible with zero and from the
Eq.(22) this condition can  be reached only if $a_P=\mid
a_{NP}\mid $ from which
\be
a_P=106~\mbox{MeV},{\rm~ and ~~
therefore}~~\overline{\alpha}_s(fit) = 0.40. \ee Note that this
$\overline{\alpha}_s$ is in agreement with expected values (26).
Correspondingly, from (23) in theory  tensor splitting
$c_P=\frac{2}{3}a_P$ is positive,
\be c = 71~\mbox{MeV}\ee Thus  positive sign of $c(1P)$ appears to
be in contradiction with the ``experimental'' number (\ref{9}):
$c_A(exp)=(-148\pm 38)$ MeV.

The second discrepancy is that spin-averaged mass in considered
case A,
\be
M_{cog} = 1303~\mbox{MeV},\ee is larger than the experimental
number (16) following from linear  Regge $L$-trajectory for
spin-averaged masses \cite{14}.

\section{The $a_0(980)$ is the $q\overline{q}$ $1~^3P_0$ state}

Here  the mass $\tilde{M}_0~(^3P_0)$ will be calculated in
one-channel approximation,
\be
\tilde{M}_0 = M_{cog}(1~^3P_J) - 2a -c \ee with $
M_{cog}(1~^3P_J)=(1260\pm 10)~\mbox{MeV}$ from experimental Regge
trajectory. We give below FS splittings $a(1P)$ and $c(1P)$ for
two values of strong coupling (24).

For $\alpha_s=0.43$ when radiative corrections of $\alpha^2_s$
order were supposed to be small, $\langle
r^{-3}\rangle_{1P}=0.0394$ GeV$^3$, one obtains ($a_p=143$ MeV,
$a_{NP}=-99$ MeV)
\be
a=44~\mbox{MeV},~~~~c=96~\mbox{MeV}\ee and from  the definition
(33)
\be \tilde{M}_0 (\alpha_s =0.43) = (1076 \pm 10)~\mbox{MeV},\ee
 so that corresponding  hadronic shift of the $a_0$ meson, according
to the definition (9), is
\be
\Delta_{had}(\alpha_s=0.43)=(91 \pm 10) ~\mbox{MeV}, \ee If
radiative corrections are not suppressed then it would be more
consistent to take in OGE terms the  strong coupling in one-loop
approximation when from Ref.[17]
\be \bar{\alpha}_s (1-loop) \cong 0.9 \alpha_{cr}(1-loop) = 0.53
\ee Then FS splittings appear to be in good agreement with the
experimental numbers
\be
a=78~\mbox{MeV},~~c=118~\mbox{MeV},\ee which correspond to the
hadronic shift equal zero and,
\be
\tilde{M}_0=(986\pm 10)~\mbox{MeV},\ee coincides with the mass of
the $a_0(985)$. Thus the hadronic shift turns out to be very
sensitive to the choice of strong coupling and:
\be
\Delta_{had}(\alpha_s=0.53)\cong (3\pm 10)~\mbox{MeV} \ee It is
essential that in Ref.\cite{9}, as well as in eqs.(37), large
value of  one-loop coupling was used. It assumes that $\alpha^2_s$
corrections to FS splitting are not small being about $(30\div$
20)\%. To understand which choice  of $\overline{\alpha}_s$ is
preferable  one needs to study many  other $P$-wave multiplets. In
any case one can conclude that predicted value of hadronic shift
is correlated with the choice of strong coupling in SO and tensor
potentials.

For FS splittings  (38) the masses of the $a_1$ and $a_2$ mesons
are equal 1242 and 1325 MeV, i.e. very close to their experimental
values.

\section{Hyperfine shift as a tool to identify the members of the $^3P_J$
multiplet}

Hyperfine (HF) splitting of the $P$-wave multiplet which comes
from spin-spin interaction, $\Delta_{HF}(1P)$, is defined as
\be
\Delta_{HF} (nP) = M_{cog} (n~^3P_J) - M(n~^1P_1)\ee where
$M_{cog}$ is not sensitive to  taken value $\tilde{M}_0$ since it
enters $M_{cog}$ with small weight equal 1/9. In (43) $M(1~^1P_1)$
is the mass of the $b_1(1235)$.

Then if the $a_0(1450)$ belongs to the lowest
$q\overline{q}~^3P_J$ multiplet, $M_{cog}=1303$ MeV and
\be
\Delta_{HF}^{(1P)} = (73\pm 3(exp)) ~\mbox{MeV},~~~~\mbox{case
A}\ee Identifying the $a_0(980)$ as the $q\overline{q}$ ground
$^3P_0$ state, one obtains $M_{cog}=1252$ MeV and
\be
\Delta_{HF}^{(1P)} = 28 \pm 6(th) \pm
3(exp)~\mbox{MeV},~~~~\mbox{case B}\ee i.e. essentially smaller.
One can compare these numbers with theoretical predictions from
Ref.[7] where   perturbative contribution to HF shift was shown to
be small and negative while NP term is positive and not small,
\be
\Delta_{HF} (th) = (30\pm 10)~\mbox{MeV}.\ee In $\Delta_{HF} (th)$
(44)  theoretical error comes from uncertainty in our knowledge of
gluonic correlation lenghth  $T_g$ in lattice calculations:
$T_g=0.2$ fm in quenched QCD and $T_g=0.3$ fm in full QCD [16].
>From comparison (44) and (43) one can see that   theoretical
number (44) appears to be in good agreement with experimental one
(43)  if the $a_0(980)$ is  the $q\overline{q}~1~^3P_0$
 and this statement does not depend on a value of hadronic shift
$\Delta_{had}$.

On the contrary if the $a_0(1450)$ is taken as the
$q\overline{q}~1~^3P_0$ state,  then ``experimental'' value (42)
is two times larger than $\Delta_{HF}(theory)$.

\section{Conclusions}

Experimental data on FS splittings in light mesons appear to be  a
useful tool to identify the  members of the $^3P_J$ multiplet. Our
study of the FS has shown that

The $a_0(1450)$ cannot be a candidate for the ground $^3P_0$ state
since under such identification\\ i)``experimental'' value of
tensor splitting turns out to be negative (with large magnitude),
$c_{exp} = (-150\pm 40)$ MeV, in contradiction with the
conventional theory. \\ ii) Also for such interpretation
spin-averaged mass $M_{cog}=1303$ MeV would be too large and lie
above linear Regge trajectory for spin-averaged masses.\\ iv)
Hyperfine shift of the $b_1(1235)$ with respect to $M_{cog}$ would
be two times larger than predicted number in Ref.\cite{7}.

There is no such discrepancies if the $a_0(980)$ is the
$q\overline{q}$ $1~^3P_0$ state. In this case the mass
$\tilde{M}_0(^3P_0)$ can be calculated in one-channel
approximation and from the difference $\tilde{M}_0(^3P_0) -
M_0(a_0(980)) =\Delta_{had}$  hadronic shift is found to be very
sensitive to the chosen  value of strong coupling
$\overline{\alpha}_s$.

If radiative corrections of $\alpha^2_s$ order are small, as in
charmonium, and $\overline{\alpha}_s \cong 0.40$ is used, then
$\Delta_{had}$ is large, $\Delta_{had}=(100\pm 10)$ MeV. If for
$\overline{\alpha}_s$ the value $\overline{\alpha}_s(1-loop)=0.53$
is taken, the  hadronic shift appears to be equal zero.

\bigskip

The author is grateful to Yu.A.Simonov for fruitful discussions.
This work has been supported by RFFI grant 00-02-17836 and INTAS
grant 00-00110.

\newpage
\vspace{1cm}

\hfill {\bf \large Appendix.}

\vspace{7mm}

\setcounter{equation}{0}
\def\theequation{A.\arabic{equation}}

 {\bf \large Spin-dependent potentials in light mesons}

\bigskip

In Refs. [12,21] all NP spin-dependent potentials in light mesons:
$\hat{V}_{LS} = {\bf L}{\bf S} V_{LS}$;~tensor potential
$\hat{V}_T=\hat{S}_{12}~V_T$, and HF potential $\hat{V}_{HF} ={\bf
S}_1{\bf S}_2 V_{HF}$ were obtained being expressed through
bilocal vacuum correlation functions (v.c.f.) $D(x)$ and $D_1(x)$
\be
V^{NP}_{LS} (r) = -\frac{1}{\mu^2_0 r} \int\limits^{\infty}_0 d\nu
\int\limits^r_0 d\lambda \Biggl (1-\frac{4\lambda}{r} \Biggr )
D(\sqrt{\lambda^2+\nu^2})
+\frac{3}{2\mu^2_0}\int\limits^{\infty}_0 d\nu
D^{NP}_1(\sqrt{r^2+\nu^2})\ee
\be
V^{NP}_T(r) = -\frac{2r^2}{3\mu^2_0}\int\limits^{\infty}_0 d\nu
\frac{\partial}{\partial r^2} D^{NP}_1(\sqrt{r^2+\nu^2})\ee
\be
V^{NP}_{HF}(r) = \frac{2}{\mu^2_0}\int\limits^{\infty}_0 d\nu
\Biggl [D(\sqrt{r^2+\nu^2}) + D^{NP}_1 (\sqrt{r^2+\nu^2})
+\frac{2r^2}{3} \frac{D^{NP}_1(\sqrt{r^2+\nu^2)}}{\partial
r^2}\Biggr ] \ee Here v.c.f. $D$ and $D_1$ are defined though the
gauge-invariant bilocal vacuum correlators:
$$\frac{g^2}{N_c} \langle F_{\mu\nu}
(x)\phi(x,y)F_{\lambda\sigma}(y) \phi(y,x) \rangle = $$
\be
(\delta_{\mu\lambda} \delta_{\nu\sigma}
-\delta_{\mu\sigma}\delta_{\nu\lambda})D(x-y) +\frac{1}{2}
\partial_{\mu}\{[~h_{\lambda}\delta_{\nu\sigma}-h_{\sigma}
\delta_{\nu\lambda}) + \mbox{permutation}~]D_1(x-y)\ee where
$h_{\mu}=x_{\mu}-y_{\mu}$ and the factor
\be
\phi(x,y) =P~exp\int\limits^x_y A_{\mu}(z) dz_{\mu}\ee
 provides the gauge invariance of the correlators (A.4). In
 Ref.[22] it was shown that in bilocal approximatiom there is no
 perturbative contribution to v.c.f. $D(x)$ while the correlator
 $D_1$ contains both perturbative and NP contributions:
\be
D_1 = D_1^{NP} + D^{\mbox{pert}}_1\ee with
\be
D_1^{\mbox{pert}}(x) = \frac{16}{3\pi}~\frac{\alpha_s}{x^4} \ee To
derive the expressions (A.1)-(A.3) the meson Green's function in
FSF representation (which is gauge -- invariant) was studied and
spin terms enter $G_M(x,y)$ through the exponential factors,
\be
exp(g\int\limits^s_0d\tau~\sigma_{\mu\nu}^{(1)} F_{\mu\nu}) exp
(-g\int\limits^{\bar{s}}_0
d\bar{\tau}\sigma_{\mu\nu}^{(2)}F_{\mu\nu})\ee Here
$\sigma_{\mu\nu} = \frac{1}{4}(\gamma_{\mu}\gamma_{\nu} -
\gamma_{\nu}\gamma_{\mu})$, $F_{\mu\nu}$ is the field strength,
and $s(\bar{s})$ is the proper time of the quark (antiquark). The
proper time $\tau(\bar{\tau})$ plays the role of ordering
parameters along the quark (antiquark) trajectory
$z(\tau)~(\bar{z}_{\mu}(\tau))$.

To obtain Hamiltonian and potentials from the meson Green's
function it is necessary to go over from the proper time to the
actual time $t$ of the quark thus defining the new quanity
$\mu(\tau)$:
\be
2\mu(t) =
\frac{dt}{d\tau},~~~2\bar{\mu}(\bar{t})=\frac{dt}{d\bar{\tau}}\ee
Then in (A.8) the integrals can be rewritten as
\be
J_q = \int\limits^s_0 d\tau~\sigma_{\mu\nu}~F_{\mu\nu} = \int^T_0
\frac{dt}{2\mu(t)}\sigma^{(1)}_{\mu\nu} F_{\mu\nu}(z(t)) \ee and
correspondingly the integral
\be
\bar{J}_q = \int\limits^T_0\frac{dt}{2\bar{\mu}(t)}
~\sigma^{(2)}_{\mu\nu}~F_{\mu\nu} \ee is defined for the
antiquark. In bilocal approximation after averaging  the exponents
(A.8) will contain the bilocal correlators, or the cumulants. To
obtain spin-dependent potentials the important approximation is
that the spin factors (A.8) are considered as a perturbation and
therefore $\mu(t)$ and $\bar{\mu}(t)$ in (A.9),(A.10) can be
changed by corresponding values $\mu_0$ and $\bar{\mu}_0$
$(\mu_0=\bar{\mu}_0)$ calculated for the unperturbed Hamiltonian
$H_R$ which is defined for a meson with spinless quark and
antiquark. Notice that in FSF representation to derive
spin-dependent potentials (in light mesons) the expansion in
inverse powers of quark mass was not used.

The derivation of the meson relativistic  Hamiltonian $H_R$ and
the definition of the constituent mass,
\be
\mu_0 = \langle \sqrt{p^2+m^2}\rangle_{nL}\ee is discussed in
details in Refs. [12-13].

The v.c.f. $D$ and $D_1$ were calculated in lattice QCD [16] where
it was shown that $D^{NP}_1$ is small with compare to $D(x)$ and
even compatible with zero in full QCD. Therefore in the potentials
(A.1)-(A.3) the terms containing $D_1^{NP}$ can be omitted, in
particular, NP contribution to tensor splitting
\be
c_{NP} = \langle V_T\rangle ~~\mbox{is compatible with zero} \ee
The perturbative contribution to SO and tensor potentials which
are defined by v.c.f. (A.7) just corresponds to OGE terms
(12),(13).

Lattice measurements has also shown that v.c.f. $D$  can be
parametrized with a good accuracy as the exponent at the distances
$x \ga 0.2$ fm, i.e.
\be
D(x) = d~exp\Biggl (-\frac{x}{T_g}\Biggr ) \ee

Then from (A.1) and (A.9)
\be
V^{NP}_{LS} (r) = -\frac{\sigma}{\mu^2_0r\pi}
\int\limits^{r/T_g}_0 t K_1(t) + \frac{4\sigma}{\pi \mu^2_0}\Biggl
[\frac{2 T_g}{r^2} - \frac{1}{T_g} K_2\Biggl (\frac{r}{T_g}\Biggr
)\Biggr ]\ee where the string tension,
\be
\sigma = 2\int\limits^{\infty}_0 d\nu ~\int\limits^{\infty}_0
d\lambda~D(\sqrt{\lambda^2+\nu^2}),\ee for the exponential form of
 $D(x)$ is
\be
\sigma=\pi dT_g^2. \ee From the expression (A.15) it can easily be
shown that
\be
V^{NP}_{LS}(r \gg T_g) \to -\frac{\sigma}{2\mu^2 r}, \ee i.e.
coincides with the Thomas precession term if $T_g$ is supposed to
be small. We shall not take into account here the positive
correction to the Thomas potential coming from second term in
(A.15) since there exist two other contributions; from the
interference of perturbative and NP  terms [23] and from Coulomb
term with correct strong coupling in background fields
$\alpha_B(r)$ which imitates linear $\sigma^* r$ potential at $r
\la 0.3$ fm [17],[18].

\end{document}